\documentclass[10pt,aps,pra,amsmath,amssymb,amsfonts,amsart,floatfix,twocolumn,superscriptaddress]{revtex4-2}

\usepackage[utf8]{inputenc} 
\usepackage[T1]{fontenc}
\usepackage{graphicx}
\usepackage{bm}    
\usepackage{amsbsy}
\usepackage[separate-uncertainty]{siunitx}
\usepackage[breaklinks=true]{hyperref}
\usepackage{hyperref}
\hypersetup{
	colorlinks=true,
	linkcolor=blue,
	citecolor=blue,
	urlcolor=blue	
}

\begin{document}
	
\title{Coherent frequency combs from electrons colliding with a laser pulse}

\author{Michael~J.~Quin}
\email{michael.quin@physics.gu.se}
\affiliation{Max-Planck-Institut f\"{u}r Kernphysik, Saupfercheckweg 1, 69117 Heidelberg, Germany}

\author{Antonino~Di~Piazza}
\email{a.dipiazza@rochester.edu}
\affiliation{Department of Physics and Astronomy, University of Rochester, Rochester, NY 14627, USA}
\affiliation{Laboratory for Laser Energetics, University of Rochester, Rochester, NY 14623, USA}
\affiliation{Max-Planck-Institut f\"{u}r Kernphysik, Saupfercheckweg 1, 69117 Heidelberg, Germany}

\author{Matteo~Tamburini}
\email{matteo.tamburini@mpi-hd.mpg.de}
\affiliation{Max-Planck-Institut f\"{u}r Kernphysik, Saupfercheckweg 1, 69117 Heidelberg, Germany}

\date{\today}

\begin{abstract}
\noindent
Highly coherent and powerful light sources capable of generating soft x-ray frequency combs are essential for high precision measurements and rigorous tests of fundamental physics. In this work, we derive the analytical conditions required for the emission of coherent radiation from an electron beam colliding with a laser pulse, modeled as a plane wave. These conditions are applied in a series of numerical simulations, where we show that a soft x-ray frequency comb can be produced if the electrons are regularly spaced and sufficiently monoenergetic. High quality beams of this kind may be produced in the near future from laser-plasma interactions or linear accelerators. Furthermore, we highlight the advantageous role of employing few-cycle laser pulses in relaxing the stringent monoenergeticity requirements for coherent emission. The conditions derived here can also be used to optimize coherent emission in other frequency ranges, such as the terahertz domain.
\end{abstract}

\maketitle

\section{Introduction}
\noindent
Optical frequency combs generated by mode-locked lasers have revolutionized laser spectroscopy, enabling unprecedented precision~\cite{udem2002, cundiff2003, picque2019, fortier2019}. Recognizing their profound importance, the 2005 Nobel Prize in Physics was awarded to John L. Hall and Theodor W. H\"{a}nsch ``for their contributions to the development of laser-based precision spectroscopy, including the optical frequency comb technique''. Optical and extreme ultraviolet (XUV) frequency combs have been instrumental in the development of atomic clocks~\cite{shvydko2023, zhang2024}, the enhancement of attosecond pulse generation~\cite{pupeza2021}, and rigorous tests of bound-state quantum electrodynamics (QED)~\cite{herrmann2009}. Extending frequency combs into the soft x-ray spectral region would increase their temporal resolution, significantly enhancing our ability for precision measurements. This advancement is critical for testing fundamental physical laws and for controlling atomic and potentially nuclear processes~\cite{hanschRMP06, HallRMP06}.

Currently, XUV frequency combs are most commonly produced via high harmonic generation~\cite{gohle2005, jones2005, cingoz2012, pupeza2013}, yet the relatively low intensity of the emitted light at high harmonic orders has prompted the exploration of alternative schemes, such as pulse shaping~\cite{cavaletto2014}. In contrast, systems which employ an electron beam, such as synchrotrons or free-electron lasers (FELs), rank among the brightest available sources of x-rays but produce temporally incoherent radiation~\cite{huang2007, schmueser2009}. Terahertz frequency combs have already been produced by electron beams, either accelerated by a synchrotron~\cite{tammaro2015} or passing over a metagrating~\cite{zhang2023}. Also, schemes exist which employ either a FEL oscillator~\cite{kim2008} or an electron beam interacting with two counterpropagating intense laser pulses~\cite{lv2022} in order to produce x-ray frequency combs. In particular, in~\cite{kim2008} it is shown that a FEL oscillator can produce large-yield x-ray pulses at high repetition rates, and in~\cite{lv2022} the envisaged x-ray source exhibits an extremely narrow relative bandwidth and a very high brilliance. 

In this paper, we focus on the collision between an electron beam and a laser pulse, modeling the laser field as a plane wave. First, we analytically investigate how the coherence of the emitted radiation depends on the initial positions and velocities of the particles, as well as the pulse duration and the angle of observation. Then, we apply these conditions to  demonstrate how a regularly-spaced beam of electrons can produce a frequency comb upon colliding with a laser pulse. A series of numerical simulations is performed to test these conditions and assess the viability of this scheme in the soft x-ray domain. 

The laser pulse considered here is well within current experimental capabilities and can be assumed not to be tightly focused, such that the plane wave approximation is appropriate, and our analytical conditions can be used for quantitative predictions. There is an extensive body of literature describing the spectrum of radiation emitted by a single electron in a plane wave; we refer the reader to the reviews~\cite{dipiazza2012, gonoskov2022, fedotov2023} and the references therein. The variation in coherence with the initial positions is relatively well understood if the initial velocities are identical~\cite{quin2020_master, gelfer2023}. However, remarkably few studies address how the coherence varies with both initial positions and initial velocities~\cite{angioi2018}, which will be the focus of our investigation here. 

The generation of both near-monoenergetic and regularly spaced electron beams is challenging. However, linear accelerators routinely produce near-monoenergetic electron beams. Microbunching, as observed in FELs, can serve as an efficient mechanism for generating a regularly spaced pattern, though it may increase the energy spread. Notably, recent advances in laser-electron beam shaping have enabled the customization of beam properties~\cite{emma2025}, and the CXFEL project aims to generate nanoscale-patterned electron beams via diffraction at similar energies to those considered here~\cite{graves2018, cxfel}. The analytical conditions derived in this study are also useful for investigating how coherent emission influences the dynamics of beams and plasmas, including systems of electrons and positrons~\cite{quin2023_prr, quin2024}. Such systems can be produced through laser-plasma interactions~\cite{sarri2015, streeterSR24, zhaoCP22}, where a regular pattern of dense electron-positron bunches may spontaneously emerge during the production process~\cite{zhaoCP22}.  

Alternatively, a laser-plasma interaction can be considered as a source of regularly spaced electrons. The analytical conditions derived in this paper are based on the plane wave approximation, where the wave phase depends on a single spatial coordinate. Within this framework, regularly spaced electrons are treated equivalently to ultrathin, regularly spaced 2D electron sheets. Notably, coherent XUV light can be generated by reflecting a laser pulse off a relativistic electron sheet, which is produced by irradiating a thin solid foil with an intense laser pulse~\cite{naumova2004, kulagin2007, wu2012, kiefer2013}. In this scenario, the requirement for near-monoenergetic electrons can be relaxed by employing a short laser pulse, as indicated by the conditions derived in section~\ref{sec:vary_energy}.

\section{Dynamics in a plane wave}
\label{sec:dynamics_planewave}

\noindent
Consider a system of $N$ electrons, each with charge $e$ and mass $m$, colliding with a laser pulse characterized by a dimensionless amplitude $a_0=|e|E_0/m\omega_0$, central frequency $\omega_0$, and a peak electric field $E_0$. Throughout this work, we use natural units, where $c=\hbar=4\pi\varepsilon_0=1$, and we take $e<0$. We model the laser field as a circularly-polarized plane wave with vector potential
\begin{equation}
	\frac{|e|}{m} \bm{A}(\varphi) = a(\varphi) \left[\bm{\hat{x}}\sin\varphi + \bm{\hat{y}}\cos\varphi\right],
	\label{eq:A}
\end{equation}
which propagates along the negative $z$ direction and satisfies the gauge condition $\nabla\cdot \bm{A}(\varphi)=0$. As we will explain below, the case of a linearly-polarized wave can be treated in a completely similar way (see also~\cite{quin2020_master}).

The pulse envelope has amplitude $a(0)=a_0$ and is assumed to vanish in the asymptotic limit $a(\pm\infty)=0$. The laser field is defined to propagate along $-z$ with wave phase $\varphi=\omega_0x_+$. Here we have introduced the light-cone coordinates for the position $x_{\pm}=t\pm z$ and velocity $u_{\pm}=\gamma\pm u_z$, which in addition to the transverse components $\bm{x}_\perp=(x, y)$ and $\bm{u}_\perp=(u_x, u_y)$ completely determine the four-vector position $x^\mu=(x_+, x_-, \bm{x}_\perp)$ and four-velocity $u^\mu=(u_+, u_-, \bm{u}_\perp)$. Throughout this paper, we employ the Minkowski metric $\text{diag}(+1, -1, -1, -1)$ and shorthand notation $(ab)=a^\mu b_\mu$ for the inner product.

Now, consider a specific electron $j$, with initial velocity $u^{j,\mu}_{0}\equiv u^{j,\mu}(\varphi^j_0)$ defined at the initial phase $\varphi^j_0=\omega_0x^j_{0,+}$. In the regime of classical electrodynamics, provided strong-field effects such as radiation reaction are negligible~\cite{dipiazza2012,gonoskov2022}, we can solve the Lorentz equation exactly for the velocity in terms of the light-cone coordinates~\cite{landaulifshitz_vol2}
\begin{align}
	\bm{u}^j_{\perp}(\varphi^j) &= -\frac{e}{m} \bm{A}(\varphi^j),
	\label{eq:u_perp}
	\\
	u^j_-(\varphi^j) &= \frac{1}{u^j_{0,+}}\left(1+[\bm{u}^j_\perp(\varphi^j)]^2\right),
	\label{eq:u_minus}
	\\
	u^j_+(\varphi^j) &= u^j_{0,+}.
	\label{eq:u_plus}
\end{align}
Here we have assumed that every electron is initialized outside the envelope $a(\varphi^j_0)=0$, where the initial phase $\varphi^j_0$ is necessarily finite. This condition can be satisfied exactly in the case of a finite-duration pulse. However, in the case of a Gaussian pulse, it is necessary to ensure that the chosen value of $\varphi^j_0$ does not significantly affect the trajectories or radiation spectrum~(see appendix~\ref{ap:general_observer}). We have also assumed that all electrons are initially counter-propagating with respect to the plane wave such that $\bm{u}^j_{0,\perp}=0$. In this case, the on-shell condition reduces to $u^j_{0,+}u^j_{0,-}=1$ at the initial phase.

Since the velocity is known, the position can then be found by integration 
\begin{equation}
	x^{j,\mu}(\varphi^j)-x^{j,\mu}_0 = \int^{\varphi^j}_{\varphi^j_0} \frac{u^{j,\mu}(\varphi)}{\omega_0u^j_{0,+}} d\varphi,
	\label{eq:x}
\end{equation}
where the wave phase and proper time $\tau^j$ are related by $d\varphi^j = \omega_0 u^j_{0,+} d\tau^j$, and we have dropped the index $j$ from any dummy variables. 

\subsection{Spectrum emitted in a plane wave}
\noindent
With the trajectories known, we can write the spectrum of radiation seen by a distant observer in the direction of the unit-vector $\bm{n}$ as~\cite[equation~(14.67)]{jackson1998} 
\begin{align}
	\frac{d\mathcal{E}}{d\omega d\Omega}& = \frac{e^2\omega^2}{4\pi^2\omega_0^2} \nonumber
	\\
	&\times \left|\sum^N_{j=1} \int_{-\infty}^{+\infty} \frac{\bm{n}\times [\bm{n}\times\bm{u}^j(\varphi^j)]}{u^j_{0,+}} e^{i\omega(nx^j(\varphi^j))} d\varphi^j \right|^2.
	\label{eq:spec_general}
\end{align}
Here $n^\mu=(1, \bm{n})$ is approximately constant at all times, for all electrons. If the observer is located in the direction along which the electrons are initially propagating $\bm{n}=\bm{\hat{z}}$, then the phase of the integral becomes $\omega x^j_{-}(\varphi^j)$, which is obtained by integrating $u^j_{-}(\varphi^j)$. Therefore, we recognize two asymptotic solutions for the spectrum; (i) $a_0\ll 1$, where the integral effectively becomes a Fourier transform of the envelope, and (ii) $a_0\gg 1$, where we can apply the stationary phase method~\cite{narozhnyi1996,kharin2016,seipt2013}. It is the first regime (i) which is usually of interest for the production of frequency combs, and which we consider here. Additionally, frequency combs can also be produced in the intermediate regime $a_0\gtrsim 1$ via the polarization grating technique~\cite{valialshchikov2021}.

After inserting the trajectory into equation~\eqref{eq:spec_general}, the spectrum of radiation emitted along $+z$ can be written as
\begin{equation}
	\frac{d\mathcal{E}}{d\omega d\Omega}\Bigg|_{\bm{n}=\bm{\hat{z}}} = \frac{e^2\omega^2}{4\pi^2\omega_0^2} \left|\sum^N_{j=1} e^{-i\omega(1 + D^j_0)z^j_{0}}\bm{I}^j(\omega)\right|^2.
	\label{eq:spec_z}
\end{equation}
For convenience, the initial conditions are defined for all particles at $t=0$. Note that the inverse Doppler shift $D^j_0=(u^j_{0,+})^{-2}$ is negligible $D^j_0\ll 1$ in the ultra-relativistic limit $\gamma^j_0\gg 1$, where $u^j_{0,+}\approx 2\gamma^j_0$. In this case, the coherence at a given frequency $\omega$ depends on the initial positions via the sum of amplitudes under the square modulus, while the dependence on the initial velocity is given by
\begin{align}
	\bm{I}^j(\omega) &= \frac{1}{u^j_{0,+}} \int_{-\infty}^{+\infty} \frac{e}{m} \bm{A}(\varphi) e^{i\phi^j(\omega, \varphi)} d\varphi,
	\label{eq:I}
	\\
	\phi^j(\omega, \varphi) &= \frac{\omega}{(u^j_{0,+})^2\omega_0} \left(\varphi + \int^{\varphi}_{-\infty}[a(\varphi')]^2d\varphi'\right).
	\label{eq:phi}
\end{align}
This Fourier-like integral characterizes the shape of the spectrum. Note that $\bm{I}^j(\omega)$ and $\phi^j(\omega, \varphi)$ are independent of the initial phase and positions provided all particles begin outside the pulse, i.e. $\bm{A}(\varphi)=\bm{0}$ for $\varphi\in[-\infty, \varphi^j_0]$. Hence, the interference pattern depends on the initial positions \textit{only} via the oscillatory terms under the square modulus. In fact, this holds for any direction of observation~(see appendix~\ref{ap:general_observer}).

If all electrons have the same initial velocity, then $\bm{I}^j(\omega)$ and $\phi^j(\omega, \varphi)$ become identical for all particles. The coherence of the emitted radiation in this case has been studied previously~\cite{quin2020_master, gelfer2023}, where equations~\eqref{eq:spec_z}--\eqref{eq:phi} are consistent with equations~(5.13)--(5.14) of~\cite{quin2020_master} and equations~(3)--(7) of~\cite{gelfer2023}.

\subsection{Variation of the initial energies}
\label{sec:vary_energy}
\noindent
In practice, to understand how the interference pattern varies with the initial velocity one must solve the integral $\bm{I}^j(\omega)$. Now, if we consider a Gaussian pulse $a(\varphi)=a_0\exp(-\varphi^2/\Delta^2)$ with a small amplitude $a_0\ll 1$ then the spectrum is given by a Fourier transform of the envelope ${\tilde{a}(k) = \int_{-\infty}^{+\infty}a(\varphi)\exp(-ik\varphi)d\varphi}$, which can be solved exactly ${\tilde{a}(k) = a_0\Delta \sqrt{\pi} \exp(-\Delta^2k^2/4)}$. In fact, there will be a peak in the positive and negative frequency domains, that is $\tilde{a}(k-1)$ and $\tilde{a}(k+1)$ respectively. 
As explained in reference~\cite{jackson1998} equations (14.59--14.60), only the positive-frequency contribution has to be considered
\begin{equation}
	\left|\bm{I}^j(\omega)\right|^2 = \frac{1}{2(u^j_{0,+})^2} \, \left[\tilde{a}\left(\frac{\omega}{(u^j_{0,+})^2\omega_0} -1\right) \right]^2.
	\label{eq:I_smalla0}
\end{equation}

Now, we can express the initial velocity of each electron as a small variation around a mean value, which can be written in terms of the initial light-cone coordinate $u^j_{0,+}=u_{0,+} + \delta u^j_{0,+}$. We intend to apply the same logic to the spectrum, by writing the contribution of each electron as a small perturbation around a mean value. Although $\bm{I}^j(\omega)$ depends explicitly on $u^j_{0,+}$, we assume the exponential dependence via $\tilde{a}(k)$ is dominant. Then, we can expand the exponent to second order in terms of the quantity $\rho^j_0=\delta u^j_{0,+}/u_{0,+}$, which is assumed to be small $|\rho^j_0|\ll 1$, such that we obtain
\begin{align}
	\tilde{a}\left(\frac{\omega}{(u^j_{0,+})^2\omega_0} -1\right) &\approx \tilde{a}(\varpi-1) \nonumber
	\\
	&\times e^{-\Delta^2\left[\varpi(1-\varpi)\rho^j_0 + \frac{1}{2}\varpi(5\varpi-3)(\rho^j_0)^2\right]}.
	\label{eq:a_tilde}
\end{align}
For an ultra-relativistic particle, $\rho^j_0$ can be more intuitively written as a variation in the initial energy or Lorentz factor $\rho^j_0\approx\delta\gamma^j_0/\gamma_0$. Here $\varpi=\omega/\omega_1$ is the frequency in units of the Doppler-shifted value $\omega_1=u^2_{0,+}\omega_0\approx4\gamma_0^2\omega_0$.

Now, consider the leading-order term in the exponent of equation~\eqref{eq:a_tilde}. This can be either positive or negative depending on the frequency observed and whether or not the particle has a higher (or lower) initial energy than the mean value. For a higher initial energy $\rho^j_0>0$, we observe an increase in the energy radiated at high frequencies $\varpi>1$ and a decrease at low frequencies $\varpi<1$, as expected. Conversely, for a lower initial energy $\rho^j_0<0$, we have an increase in the energy radiated for $\varpi<1$ and a decrease for $\varpi>1$.
Note that these are perturbative changes on top of the underlying spectrum $\tilde{a}(\varpi -1)$, so one can never observe indefinite exponential growth of the energy radiated at a given frequency.

If we observe at the peak of the spectrum $\varpi=1$, then the leading-order term vanishes and we must consider the second-order term in the exponent of equation~\eqref{eq:a_tilde}. At this frequency, any deviation in any particle's initial energy from the mean value results in reduced radiated energy and diminished coherence. For the emission to be coherent at $\varpi=1$, equation~\eqref{eq:a_tilde} indicates that the exponent should be significantly smaller than unity. Therefore, the condition for coherence on the energy is $(\rho^j_0\Delta)^2\ll 1$. This implies that the standard deviation of the initial energy $\sigma_{\gamma_0}$ of the beam must satisfy $\sigma_{\gamma_0}\Delta \ll \gamma_0$.

In other words, a short pulse and small energy spread are optimal for producing coherent radiation at the peak of the spectrum. Note that the parameter $\Delta$ is simply related to the time interval required for the interaction with the laser pulse to take place. If we repeat this derivation for a linearly-polarized pulse, e.g. $\bm{A}(\varphi)=a(\varphi)\cos\varphi\hat{\bm{x}}$, we find that the condition for coherence on the velocities is unchanged with the replacement $a_0^2\to a_0^2/2$ (in the regime $a_0\ll 1$). 

\subsection{Variation of the initial positions}
\noindent
Now, we argue that if the above condition on the initial energy spread is satisfied, then we can effectively treat all electrons as if they have the same initial velocity. In this case $\bm{I}^j(\omega)$ and $\phi^j(\omega, \varphi)$ are identical for all electrons, and the coherence will be determined by the sum of oscillatory terms under the square modulus in equation~\eqref{eq:spec_z}. Coherent emission will then occur at a given frequency $\omega$ when the phase difference is an integer multiple of $2\pi$. An analogous condition also applies to a general angle of observation, as discussed in appendix~\ref{ap:general_observer} and~\cite{quin2020_master, gelfer2023}. 

If we consider many electrons $N\gg 1$, which are ultra-relativistic and regularly spaced along the $z$-axis at $z^j_0=z_0-jd$, then equation~\eqref{eq:spec_z} becomes the well-known definition of a frequency comb~(see e.g.~\cite{fortier2019}), where the spacing of the harmonics is $\Delta\omega=2\pi/d$. In the regime $a_0\ll 1$, the spectrum is characterized by a peak around the Doppler-shifted frequency $\omega_1$ with full-width-at-half-maximum $\text{FWHM}_\omega=2\omega_1\sqrt{2\ln 2}/\Delta$. Hence, we also require $\Delta\omega\ll\text{FWHM}_\omega$ to observe several harmonics inside the bandwidth. From now on, we will refer to this electron beam, with regular spacing and identical initial velocities, as the `ideal beam'.

To understand how the coherence depends on the initial positions, we perturb around the ideal case $z^j_0=z_0-jd+\delta z^j_0$. Let us assume that $d$ is chosen such that the emission is exactly coherent at a given frequency, i.e. $\omega d=2\pi l$, for integer $l$. The interference pattern in equation~\eqref{eq:spec_z} then becomes $|\sum^N_{j=1}\exp(i\omega\delta z^j_0)|^2$. If the perturbation is small $\delta z^j_0 \ll d$, then the residual phase will also be small $\omega\delta z^j_0\ll 2\pi$. A quantitative condition $\omega\delta z^j_0< \pi/5$ is given in~\cite{angioi2018}, and so we require that $\delta z^j_0 / \lambda < 0.1$ is necessary for coherent emission at wavelength $\lambda=2\pi/\omega$. In particular, we are interested in coherence near the peak of the spectrum, i.e. at $\lambda=\lambda_1$, where $\lambda_1=2\pi/\omega_1$ is the Doppler-shifted wavelength of the first harmonic. For many particles, when perturbing around the ideal case, this implies that the standard deviation $\sigma_z$ of the initial $z$ coordinate should satisfy $\sigma_z < 0.1\lambda_1$ for coherent emission to occur at the peak.

\section{Simulations}
\label{sec:simulations}
\noindent
To demonstrate how a frequency comb could be produced under the conditions described above, we have carried out a series of numerical simulations. In our code~\cite{quin2023_phd}, the trajectories are obtained by numerically integrating the Lorentz equation with a second-order leapfrog scheme~(see equation~(A.17) of~\cite{tamburini2011_phd}). With the trajectories known, the radiation spectrum in equation~(14.67) of~\cite{jackson1998} is then evaluated via a fast Fourier transform. 

Specifically, we consider an electron beam colliding with a laser pulse, modeled as a circularly polarized plane wave pulse~(see equation~\eqref{eq:A}). The laser pulse has a Gaussian envelope $a(\varphi)=a_0 \exp(-\varphi^2/\Delta^2)$, amplitude $a_0=0.1$, central wavelength $\lambda_0=800\,\si{\nano\metre}$ and pulse width $\text{FWHM}_t=15\,\si{\femto\second}$, corresponding to $\Delta=\omega_0\text{FWHM}_t/\sqrt{2\ln 2}\approx 30$. To begin with, we utilize an idealized electron beam with $N=100$ particles propagating along $+z$ with initial Lorentz factor $\gamma_0=20$. The electrons are regularly spaced at $z^j_0=z_0-jd$ with separation $d=100\lambda_1$ at time $t=0$. Here $z_0=-2.5\,\text{FWHM}_t$ is chosen such that the electrons effectively begin outside the envelope. As the electron beam is ultra-relativistic and of low density, we neglect interparticle fields and solve for the trajectories using the external field alone~\cite{quin2024, quin2023_prr}. Finally, we chose a small time step $dt=\pi/3\omega_1\approx2.8\times10^{-19}\,\si{\second}$ to fully resolve the spectrum to a frequency well above the Doppler-shifted value.

In the regime $a_0\ll 1$, most of the radiation is emitted into a cone of angle $\sim 1/\gamma_0$ around the $z$ axis. Yet, the frequency comb derived earlier was observed \textit{exactly} along the $z$-axis. In reality, one observes the radiation emitted into a range of angles. However, we can always place the detector arbitrarily far away from the interaction region such that the radiation propagating along $+z$ dominates the observed spectrum. Therefore, we choose to integrate the spectrum over the range of solid angles subtended by a cone of half-angle $0.1/\gamma_0$ around the $z$-axis, which corresponds roughly to a $2\,\si{\centi\metre}^2$ detector placed at distance $1\,\si{\metre}$.

The frequency comb emitted by the idealized electron beam can be seen in figure~\ref{fig:ideal}. Note that the spectrum is centered on the Doppler-shifted frequency $\omega_1\approx 2480\,\si{\electronvolt}$ with bandwidth $\text{FWHM}_\omega\approx 195\,\si{\electronvolt}$, and the spacing between the harmonics is $\Delta\omega\approx 25\,\si{\electronvolt}$. The dashed line is simply $N^2$ multiplied by the spectrum emitted by a single particle. If we repeat this simulation with the linearly-polarized laser pulse described at the end of section~\ref{sec:vary_energy}, the height of the frequency comb is reduced by a constant factor related to the reduced cycle-averaged field amplitude. We will now repeat this simulation while perturbing the initial positions and velocities of the electrons, to see how resilient this spectrum is, and to verify the conditions derived above.

\subsection{Variation of initial positions}
\noindent
Now, we begin from the idealized electron beam and perturb the initial $z$ coordinates according to $z^j_0=z_0-jd+\delta z^j_0$. Here, $\delta z^j_0$ is obtained from a Gaussian distribution of standard deviation $\sigma_z=[0.05, 0.10, 0.15]\,\lambda_1$ centered on $z_0-jd$ for each particle. After repeating our simulations with these parameters, one can see in figure~\ref{fig:vary_xyz} how the coherence of each harmonic declines as $\sigma_z$ increases. Recall that our definition for coherence on the initial positions was $\sigma_z/\lambda_1<0.1$ at the peak of the spectrum. This provides reasonable agreement with figure~\ref{fig:vary_xyz}, though the transition from coherent to incoherent is gradual and so the exact cut-off is arbitrary. Note that a fully incoherent spectrum should roughly correspond to the spectrum in figure~\ref{fig:vary_xyz} divided by the number of particles $N=100$.

We have carried out similar simulations while keeping the $z^j_0$ coordinates the same as in the ideal case and perturbing the initial transverse coordinates by $\sigma_x=\sigma_y=\lambda_1$. However, we found that this has virtually no effect on the spectrum, as presented in figure~\ref{fig:ideal}. Apparently this results from the fact that we only consider the radiation emitted into a narrow range of angles around the $z$-axis, such that the spectrum closely resembles equations~\eqref{eq:spec_z}--\eqref{eq:I_smalla0}. Naturally, if the spectrum is particularly sensitive to $z$ and insensitive to $x$ and $y$, then we expect a similar dependence on the respective components of the velocity, which we will now investigate.

\subsection{Variation of initial velocities}

\begin{figure}
	\centering
	\includegraphics[width=0.95\linewidth]{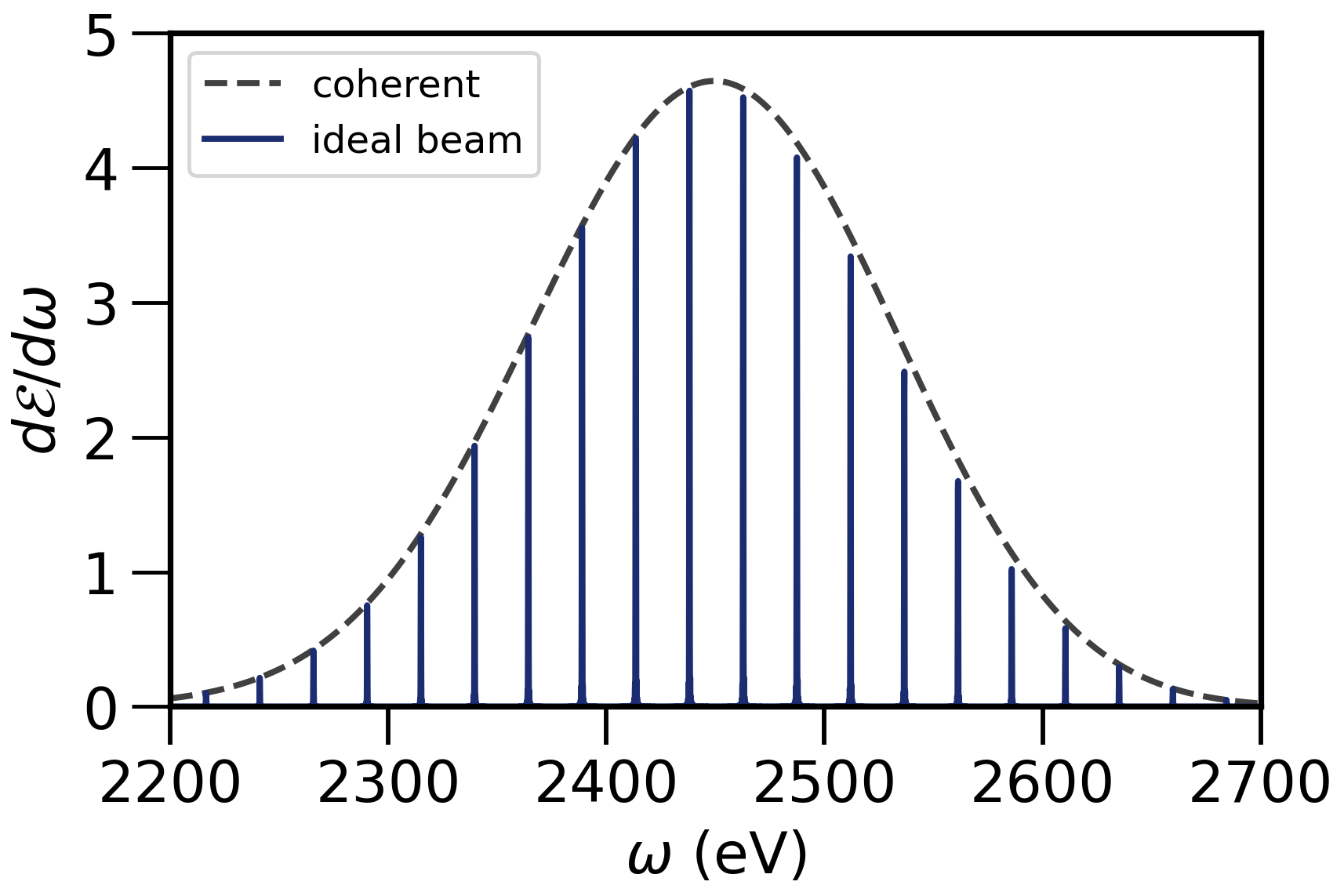}
	\caption{Frequency comb produced by an ideal electron beam colliding with a plane wave pulse. The electrons have identical initial velocities and are regularly spaced along the $z$-axis. The dashed line represents fully coherent emission. The $y$-axis shows a dimensionless quantity, as $\hbar = 1$.}
	\label{fig:ideal}
\end{figure}

\begin{figure}
	\centering
	\includegraphics[width=0.95\linewidth]{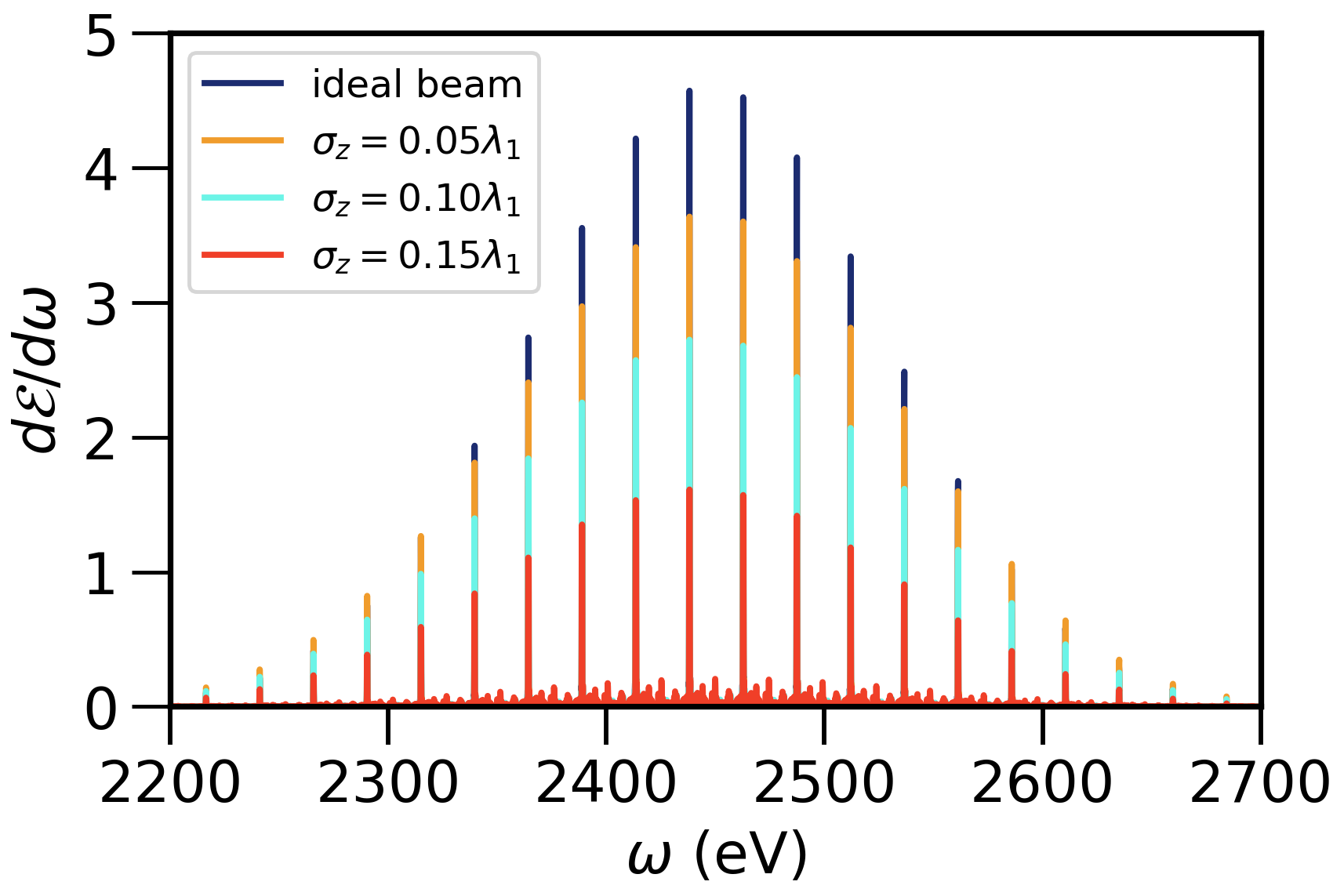}
	\caption{Dependence of the frequency comb on the initial positions of the electrons, where the initial velocities are identical. The coherence decreases as the positions are perturbed around the regular-spacing of the ideal case. Note that the harmonics near the peak are particularly strongly affected. The $y$-axis shows a dimensionless quantity, as $\hbar = 1$.}
	\label{fig:vary_xyz}
\end{figure}

\begin{figure}
    \centering
    \includegraphics[width=0.95\linewidth]{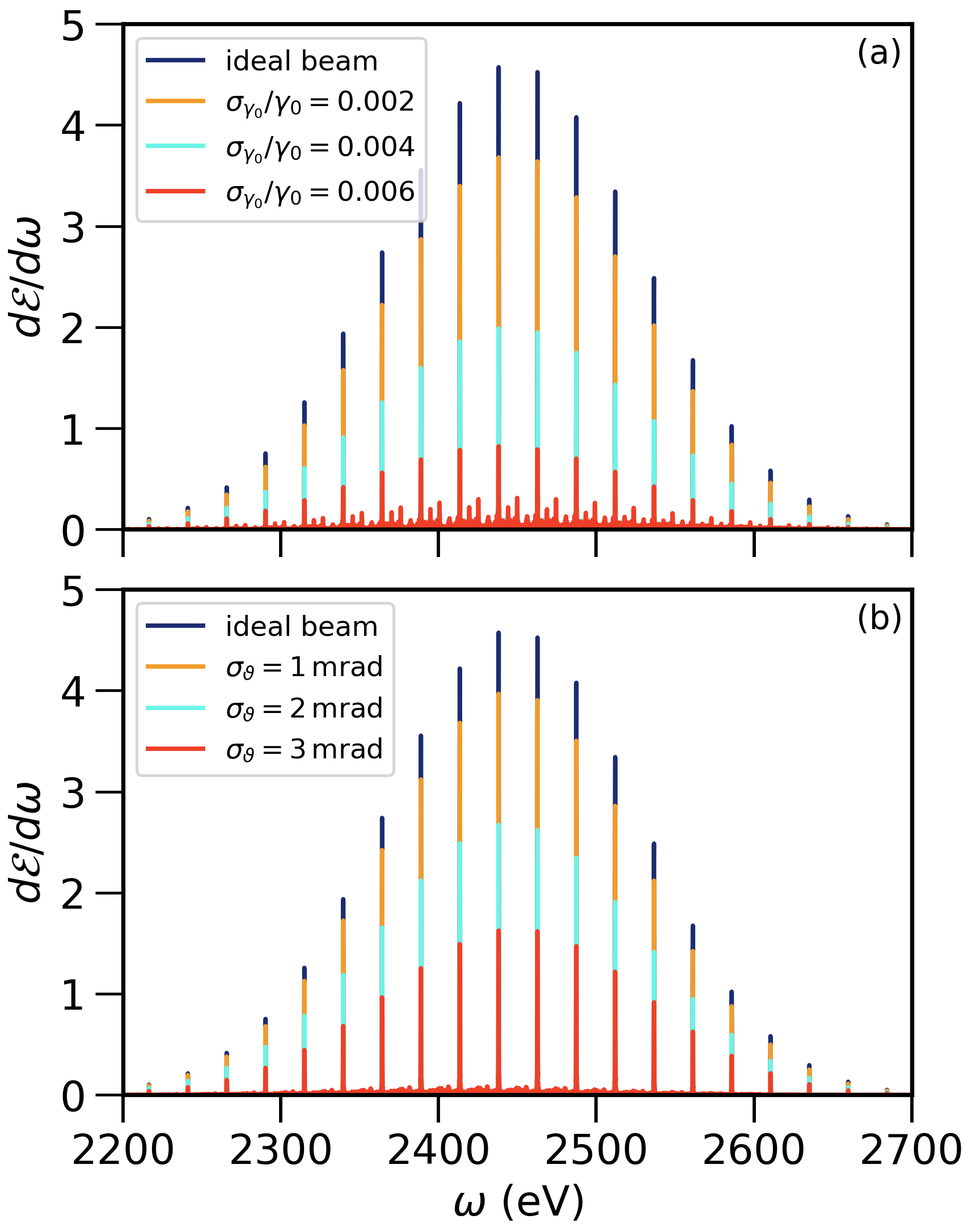}
    \caption{Dependence of the frequency comb on the initial (a) energy spread and (b) angular divergence of the electron beam. As in the ideal case, the electrons are initially regularly-spaced along the $z$-axis. If the energy spread or divergence is too large, this will decrease the coherence at each harmonic. The $y$-axis shows a dimensionless quantity, as $\hbar = 1$.}
    \label{fig:vary_u}
\end{figure}

\noindent
Once again we begin from the ideal electron beam and perturb the initial velocities. Since a spread in the initial velocities induces a change in the relative positions of the electrons also without external field, one has to make sure that the electron beam enters the laser field in a relatively short time after the initialization. 

For the sake of clarity, we first investigate the effect of the energy spread, at constant velocity direction, and then the effect of the angular spread in the velocities at constant energy. We start by varying the initial Lorentz factor of each particle with a Gaussian spread $\sigma_{\gamma_0}/\gamma_0=[2, 4, 6]\times 10^{-3}$ around the mean value $\gamma_0$. This variation in the initial energy only affects the longitudinal velocity, while the initial transverse velocity remains exactly zero. In figure~\ref{fig:vary_u}\,(a), one can see the coherence decreases with increasing the energy spread. This is in agreement with the condition derived earlier for coherence $\sigma_{\gamma_0}/\gamma_0\ll 1/\Delta\approx0.03$, and in fact, incoherence is already reached around $\sigma_{\gamma_0}/\gamma_0\sim 0.01$.

Alternatively, we can keep the initial Lorentz factor of all particles the same while introducing an angular spread (i.e., divergence) of $\sigma_\vartheta=[1, 2, 3]\,\si{\milli\radian}$ around the $z$-axis. In figure~\ref{fig:vary_u}\,(b), one can see this adversely affects the coherence at each harmonic. Although we have no analytical description for this case, we notice that $\sigma_{\gamma_0}/\gamma_0=0.002$ in figure~\ref{fig:vary_u}\,(a) and  $\sigma_{\vartheta}=2\,\si{\milli\radian}$ in figure~\ref{fig:vary_u}\,(b) are remarkably similar. This could be because the longitudinal component of the velocity is similar in both cases, i.e. $\sigma_\vartheta\gamma_0 = \sigma_{\gamma_0}$. This suggests that, although the initial transverse velocity is now non-zero, it has little impact on the spectrum emitted; once again, we are left with the conclusion that the spectrum is particularly sensitive to longitudinal positions and velocities, but has a far weaker dependence on the transverse positions and velocities.

So far in our simulations, we have varied each parameter independently to understand how it affects the radiation spectrum. However, in a realistic scenario all parameters should be varied simultaneously. Therefore, we have performed a final simulation where the initial positions are perturbed by $\sigma_z=0.02\lambda_1$ and $\sigma_x=\sigma_y=\lambda_1$ around the regular-spacing of the ideal case. The initial energy spread and angular divergence, instead, are $\sigma_{\gamma_0}/\gamma_0=0.001$ and $\sigma_\vartheta=1\,\si{\milli\radian}$, respectively. The resulting frequency comb would be indistinguishable by eye to that shown for $\sigma_z=0.1\lambda_1$ in figure~\ref{fig:vary_xyz}, or in other words, the amplitude of the harmonics near the peak is about half that of the ideal case. Also in this case, if we repeat this simulation with the linearly-polarized pulse described at the end of section~\ref{sec:vary_energy} we find the spectrum is unchanged except by a constant factor in the height (as seen for the ideal case in figure~\ref{fig:ideal}). This verifies that our conditions for coherence, which were derived in the regime $a_0\ll 1$, are essentially polarization-independent.

\section{Conclusion}
\noindent
In summary, we have derived the conditions necessary for the emission of coherent radiation from an electron beam colliding with a plane wave pulse and applied these findings to the generation of coherent frequency combs. These results can be used to demonstrate how coherence varies with the initial energy spread, particle positions, and pulse duration. Specifically, we conducted numerical simulations to explore how a soft x-ray frequency comb could be produced under these conditions. Now, the optical laser considered here is widely available. An electron beam with a small initial energy spread and angular divergence, as utilized here, can be generated from a linear accelerator. It is challenging, however, to produce thin bunches or sheets of electrons which are simultanouesly near-monoenergetic and regularly-spaced on the nanoscale, though recent developments towards this goal have been discussed in the introduction. Therefore, we note that the analytical conditions derived here could be used to further optimize this setup. For example, one could tolerate a larger initial energy spread with a few-cycle laser pulse. Alternatively, a lower frequency regime (e.g., optical or Terahertz) could be considered, allowing for more relaxed electron beam requirements.

\appendix

\section{Radiation spectrum observed in an arbitrary direction}
\label{ap:general_observer}
\noindent
Here, we show how the coherence of the emitted radiation varies with the direction of observation. This is simply a generalization of equations~\eqref{eq:spec_z}--\eqref{eq:phi}. As before, the particles collide with a plane-wave pulse $A^\mu(\varphi)=(0,\bm{A}(\varphi))$ propagating in the direction of unit-vector $\bm{n}_0$, where $n^\mu_0=(1,\bm{n}_0)$. The velocity can then be written as function of the wave phase $\varphi^j=\omega_0(n_0x^j)$ as~\cite{landaulifshitz_vol2}
\begin{align}
	u^{j,\mu}(\varphi^j) = u^{j,\mu}_0 &- \frac{e}{m}A^\mu(\varphi^j) + \frac{e(u^j_0 A(\varphi^j))}{m (n_0u^j_0)} n^\mu_0 \nonumber
	\\
	&- \frac{e^2\left(A(\varphi^j)\right)^2}{2m^2 (n_0u^j_0)} n^\mu_0.
	\label{eq:u_planewave}
\end{align}
Once again, we have assumed that all particles start outside the pulse $A^\mu(\varphi^j_0)=0$ at the initial phase $\varphi^j_0=\omega_0(n_0x^j_0)$. The position of each particle is then given by the integral in equation~\eqref{eq:x}, and according to equation~\eqref{eq:spec_general} the radiation spectrum is
\begin{equation}
	\frac{d\mathcal{E}}{d\omega d\Omega} = \frac{e^2\omega^2}{4\pi^2\omega_0^2} \left|\sum^N_{j=1} e^{i\omega(\tilde{n}^jx^j_0)} \bm{\mathcal{I}}^j(\omega, \bm{n}) \right|^2.
	\label{eq:spec_n_tilde}
\end{equation}
As in equation~\eqref{eq:spec_z}, the interference pattern depends on the initial positions via the sum of oscillatory terms under the square modulus. However, for a generic observation direction we also have a dependence on the angle of observation via the four-dimensional quantity
\begin{equation}
	\tilde{n}^{j,\mu} = n^\mu - \frac{(nu^j_0)}{(n_0u^j_0)}n^\mu_0,
	\label{eq:n_tilde}
\end{equation}
where $n^\mu=(1, \bm{n})$. The spectrum emitted by a single particle is characterized by the following integral
\begin{align}
	\bm{\mathcal{I}}^j(\omega, \bm{n}) &= \int^{+\infty}_{-\infty} \frac{\bm{n}\times(\bm{n}\times\bm{u}^j(\varphi))}{(n_0u^j_0)} e^{i\Phi^j(\omega, \bm{n}, \varphi)} d\varphi,
	\label{eq:I_cal}
	\\
	\Phi^j(\omega, \bm{n}, \varphi) &= \frac{\omega}{\omega_0}\frac{(nu^j_0)}{(n_0u^j_0)}\varphi + \frac{\omega}{\omega_0}\int^{\varphi}_{-\infty} \frac{(n\Delta u^j(\varphi'))}{(n_0u^j_0)} d\varphi'.
	\label{eq:Phi}
\end{align}

Here $\Delta u^{j,\mu}(\varphi^j)=u^{j,\mu}(\varphi^j)-u^{j,\mu}_0$ depends on the wave phase only via the vector potential. Therefore, $\Phi^j(\omega, \bm{n}, \varphi)$ is independent of the initial phase (and position) provided each particle begins outside the pulse, i.e. $A^\mu(\varphi)=0$ for $\varphi\in[-\infty, \varphi^j_0]$. Hence, the interference pattern depends on the initial positions only via the sum of oscillatory terms under the square modulus in equation~\eqref{eq:spec_n_tilde}. If all particles have the same initial velocity then $\tilde{n}^{j,\mu}$, $\bm{\mathcal{I}}^j(\omega, \bm{n})$ and $\Phi^j(\omega, \bm{n}, \varphi)$ are identical for all particles. In this case, coherent emission occurs if the phase difference in equation~\eqref{eq:spec_n_tilde} is an integer multiple of $2\pi$, as shown in equation~(5.15) of~\cite{quin2020_master}.

In the case of a finite-duration pulse, e.g. $a(\varphi)=a_0\cos^2(\varphi/\Delta)$ which is non-zero for $\varphi\in[-\pi\Delta/2, +\pi\Delta/2]$, the condition $A^\mu(\varphi^j_0)=0$ can be satisfied exactly provided $\varphi^j_0\leq-\pi\Delta/2$. In the case of a Gaussian pulse, as considered in this paper, this condition can only be satisfied approximately $A^\mu(\varphi^j_0)\approx 0$. In practice, we found via numerical simulations that if the initial phase satisfied $\varphi^j_0 \leq -2.5\,\text{FWHM}_t\,\omega_0$ then the spectra presented in this paper were not visibly affected.
Finally, we note that if the plane wave propagates in the opposite direction to the observer $\bm{n}=-\bm{n}_0$, and the initial transverse velocities are zero, then equations~\eqref{eq:spec_n_tilde}--\eqref{eq:Phi} reduce to equations~\eqref{eq:spec_z}--\eqref{eq:phi}.

\begin{acknowledgments}
\noindent
The authors wish to thank Christoph~H.~Keitel and Giuseppe Sansone for helpful discussions and for providing useful comments on the manuscript. M. J. Q. and A. D. P. wish to thank Evgeny Gelfer for discussions about coherence. M. J. Q. also thanks Chunhai Lyu for advice about the generation of optical frequency combs.

This material is based upon work supported by the Department of Energy [National Nuclear Security Administration] University of Rochester ``National Inertial Confinement Fusion Program'' under Award Number(s) DE-NA0004144.

This report was prepared as an account of work sponsored by an agency of the United States Government. Neither the United States Government nor any agency thereof, nor any of their employees, makes any warranty, express or implied, or assumes any legal liability or responsibility for the accuracy, completeness, or usefulness of any information, apparatus, product, or process disclosed, or represents that its use would not infringe privately owned rights. Reference herein to any specific commercial product, process, or service by trade name, trademark, manufacturer, or otherwise does not necessarily constitute or imply its endorsement, recommendation, or favoring by the United States Government or any agency thereof. The views and opinions of authors expressed herein do not necessarily state or reflect those of the United States Government or any agency thereof.
\end{acknowledgments}

\bibliography{bibliography}

\end{document}